\documentclass[journal,comsoc]{IEEEtran}
\usepackage{cite}
\usepackage{graphicx}
\usepackage{amsmath}
\usepackage{times}
\usepackage{latexsym}
\usepackage{graphicx}
\usepackage{bm}
\usepackage{amssymb}
\usepackage[center]{caption2}
\usepackage{stfloats}
\usepackage{cases}
\usepackage{subfigure}
\usepackage{array}
\usepackage{setspace}
\usepackage{fancyhdr}
\usepackage{citesort}
\usepackage{colortbl}
\usepackage{CJK}
\usepackage{tikz}
\usepackage{float}
\usepackage{color}

\usepackage{bm}
\usepackage{mathrsfs}

\renewcommand{\QED}{\QEDopen}

\allowdisplaybreaks[4]
\renewcommand{\baselinestretch}{0.97}

\newcaptionstyle{mystyle2}{%
\captionlabel $.$ \, \captiontext \par} \captionstyle{mystyle2}
\renewcommand\thesubsection{\Alph{subsection}}

\begin{document}
%\begin{CJK*}{GBK}{song}
\title{Deep Learning for Spectrum Sensing}
\author{\authorblockN{Jiabao Gao, Xuemei Yi, Caijun Zhong, Xiaoming Chen, and Zhaoyang Zhang}
%\thanks{Manuscript received July 4, 2019, revised September 1, 2019, accepted September 1, 2019. This work was supported in part by the NSFC-Zhejiang Joint Fund for the Integration of Industrialization and Informatization No. U1709219, the National Natural Science Foundation of China under Grant 61922071, 61871344 and 61725104. The editor coordinating the review of this paper and approving it for publication was P. Diamantoulakis. ({\it Corresponding author: Caijun Zhong})}
\thanks{J. Gao, X. Yi, C. Zhong, X. Chen, and Z. Zhang are with the College of Information Science and Electronic Engineering, Zhejiang University, Hangzhou, China, and also with the Zhejiang Provincial Key Laboratory of Information Processing, Communication and Networking, Hangzhou. (email: caijunzhong@zju.edu.cn).}
}
\maketitle

\begin{abstract}
In cognitive radio systems, the ability to accurately detect primary user's signal is essential to secondary user in order to utilize idle licensed spectrum. Conventional energy detector is a good choice for blind signal detection, while it suffers from the well-known SNR-wall due to noise uncertainty. In this letter, we firstly propose a deep learning based signal detector which exploits the underlying structural information of the modulated signals, and is shown to achieve the state of the art detection performance, requiring no prior knowledge about channel state information or background noise. In addition, the impacts of modulation scheme and sample length on performance are investigated. Finally, a deep learning based cooperative detection system is proposed, which is shown to provide substantial performance gain over conventional cooperative sensing methods.
\end{abstract}

\begin{keywords}
Spectrum sensing, SNR-wall, deep learning, cooperative detection
\end{keywords}

\section{INTRODUCTION}
Cognitive radio, which allows unlicensed devices to opportunistically utilize the licensed spectrum such as TV broadcast bands, has been proposed as a potential method to address the spectrum shortage issue \cite{CR1,CR2,CR3}. One of the key challenges for the practical deployment of cognitive radio systems is to provide sufficient protection to the licensed users. Hence, reliable detection of the presence of primary signals, which are usually very weak, is of paramount importance \cite{tandra2008snr}.

Energy detector is a widely used conventional detector due to its simplicity. However, the performance of energy detector hinges heavily on the knowledge of noise density. In practice, due to the existence of noise uncertainty, energy detector fails to work when the signal to noise ratio (SNR) falls below some threshold, commonly known as the SNR-wall. According to the existing literature, the SNR-wall for practical noise uncertainty is about $-6$ dB, which is far away from the SNR limit of $-15$ dB as required by IEEE 802.22. In\cite{tandra2008snr}, the authors suggested three different approaches to get around the SNR-wall, namely, exploiting the structure of the primary signal, using diversity and reducing the noise uncertainty.

Since the secondary users often do not have any prior knowledge of the primary signals, it is desirable to devise a blind sensing method, which can identify the underlying structure of the primary signals. Recently, deep learning (DL) has demonstrated its remarkable potential in extracting the hidden structure of different objects in various complicated tasks such as computer vision \cite{krizhevsky2012imagenet} and wireless communication \cite{jiang2018artificial}. Comprehensive reviews of the application of DL in the physical layer can be found in \cite{o2017introduction} and \cite{wang2017deep}. In the context of spectrum sensing, machine learning approaches have also been proposed in the literature \cite{Azmat,DLSS}. In particular, \cite{DLSS} proposed a DL based spectrum sensing method for OFDM systems, where a stacked autoencoder is used for feature extraction.

Motivated by the encouraging results of \cite{DLSS}, in this letter, we firstly propose a DL based detector using convolutional long short-term deep neural networks (CLDNN)\cite{cldnn}, which is applicable for arbitrary types of primary signals. It is worth highlighting that the proposed detector does not require any additional information of the primary signal or noise density when deployed online. Moreover, to further improve the sensing performance, a DL based soft combination strategy is proposed for cooperative detection. According to the simulation results, the proposed DL based detection methods significantly outperform the conventional methods.

%\textcolor{blue}{$\bullet$
%It proposes a DL based signal detector with state of the art performance, and thoroughly investigates the impacts of modulation scheme and sample length on detection performance.}
%
%\textcolor{blue}{$\bullet$
%It proposes a two-stage training strategy to deal with metrics inconsistency between general DL algorithms and specific communication tasks, and achieves relatively precise detection performance control.}
%
%\textcolor{blue}{$\bullet$
%It proposes a DL based cooperative detection system, which can simultaneously make good use of spectrum holes and provide sufficient protection to the primary user.}

\section{Problem formulation}
Depending on idle or busy state of the primary user, the signal detection at the secondary user can be modeled as the following binary hypothesis testing problem\cite{liang2008sensing,quan2009optimal}

\begin{equation}
\label{ProblemFormulation}
\bm{y}(n)=
\left\{
             \begin{array}{lr}
             \bm{w}(n) & :\mathscr{H}_0\\
             \bm{hs}(n)+\bm{w}(n) & :\mathscr{H}_1
             \end{array}
\right.,
\end{equation}
where $\bm{y}(n)$ is the $n$-th received sample, $\bm{s}(n)$ is the signal from the primary user, $\bm{h}$ is channel gain which is assumed to remain unchanged during the sensing period \cite{ED}, and $\bm{w}(n)$ is additive noise following the zero mean circularly symmetric complex Gaussian (CSCG) distribution with variance $2{\sigma_w}^2$. Also, $\mathscr{H}_0$ and $\mathscr{H}_1$ are the two hypotheses denoting the absence and presence of primary signal in a certain band, respectively.

For the conventional energy detector, the test statistic is the energy of the received signal normalized with respect to the sample number $N$ and noise variance $2{\sigma_w}^2$, as given by\cite{mariani2011effects}
\begin{equation}
\label{statistic}
\Lambda = \frac{1}{2{\sigma_w}^2N}\sum_{n=1}^N{\vert \bm{y}(n)\vert^2}.
\end{equation}
Hence, the two key performance measures for the energy detector, namely, false alarm probability and missed detection probability can be respectively expressed as $P_f={\sf Pr}(\Lambda>\lambda \vert \mathscr{H}_0)$ and $P_{md}={\sf Pr}(\Lambda<\lambda \vert \mathscr{H}_1)$, where $\lambda$ denotes SNR threshold. Also, the probability of detection is given by $P_{d}=1-P_{md}$. A good detector needs to achieve both low $P_f$ and $P_{md}$ (less than $10\%$), even at very low SNRs.

In practice, the noise density can be estimated by using noise only samples, and the SNR-wall under certain performance requirement is given by \cite{mariani2011effects}
\begin{equation}
\label{SNR-wall}
\gamma_{min} = \frac{1-Q^{-1}(P_d)\sqrt{\phi}}{1-Q^{-1}(P_f)\sqrt{\phi}}-1,
\end{equation}
where $N$ is the sample number for detection, $M$ is the number of noise only samples, $\phi$ is $\frac{N+M}{NM}$, and $Q^{-1}(x)$ is the inverse of Gaussian Q-function.

\section{DL BASED DETECTOR}
\label{system}
We now introduce the DL based sensing framework. In general, the detection algorithm $\bm{D_{dl}}$ can be expressed as
\begin{equation}
\label{Ddeep}
\bm{D_{dl}}(\bm{y})= {\sf argmax}(f^{L}(f^{L-1}(f^{L-2}(\cdots f^1(\bm{y}))))),
\end{equation}
where the input $\bm{y}$ is the vector of received samples, which is processed through a customized neural network consisting of $L$ layers. $f^i$, $i=1, \cdots, L-1$, represents for the computation with weights and
activation function of the $i$-th layer. $f^L$ is the SOFTMAX function which gives the probabilities of two hypotheses, and ${\sf argmax}$ is an operator returning the index of the largest number in a list.

\subsection{Network Architecture Design}
Inspired by the result of \cite{mr} where CLDNN performs best in modulation recognition tasks, we also adopt this kind of architecture in this letter. The superiority of CLDNN over other popular neural network architectures will be validated through numerical simulations as well.

\begin{figure}[htbp]
\centering
\includegraphics[width=0.4\textwidth]{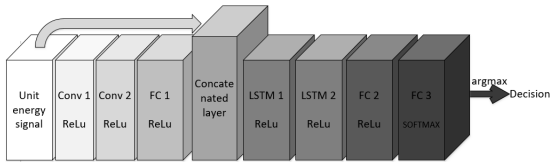}
\caption{Network architecture of DetectNet.}
\label{cldnn}
\end{figure}

It turns out that a network with two convolution (Conv) layers, two long short-term memory (LSTM) layers, one fully-connected (FC) layer after Conv layers and two FC layers after LSTM layers yields best performance. For activation function, FC3 uses SOFTMAX while all other layers use ReLu. Dropout is also used after every layer to prevent overfitting. The above model is termed as ``DetectNet'' and its network architecture is illustrated in Fig. \ref{cldnn}. Hyperparameters determined through extensive cross-validation are detailed in Table \ref{hyperparameter}.
\begin{table}[htbp]
\begin{tabular}{|c|c|}
\hline
Hyperparameter & Value \\ \hline
Filters per Conv layer & 60 \\ \hline
Filter size & 10 \\ \hline
Cells per LSTM layer & 128 \\ \hline
Neurons per FC layer & 128 $\&$ Sample length $\&$ 2 \\ \hline
Optimizer & Adam \\ \hline
Initial learning rate & 0.0003 \\ \hline
Batch size & 200 \\ \hline
Dropout ratio & 0.2 \\ \hline
\end{tabular}
\centering
\caption{Hyperparameters of the proposed CLDNN}
\label{hyperparameter}
\end{table}

\subsection{Dataset Generation and Preprocessing}
For dataset, we generate 8 kinds of digitally modulated signals at different SNRs as positive samples as per RadioML2016.10a \cite{radioml}, which is a widely used baseline dataset in modulation recognition tasks, and the negative samples are CSCG noises. The entire dataset is partitioned into three different sets for training, validation and testing with a commonly used split ratio of 3:1:1. Dataset parameters are detailed in Table \ref{dataset}.

\begin{table}[htbp]
\newcommand{\tabincell}[2]{\begin{tabular}{@{}#1@{}}#2\end{tabular}}
\begin{tabular}{|c|c|}
\hline
Modulation scheme & \tabincell{c}{BPSK,QPSK,8PSK,CPFSK\\QAM16,QAM64,GFSK,PAM4} \\ \hline
Samples per symbol & 8 \\ \hline
Sample length & 64, 128, 256, 512, 1024 \\ \hline
SNR range & -20$\sim$20dB in 1-dB increments \\ \hline
Training samples & 48000 \\ \hline
Validation samples & 16000 \\ \hline
Testing samples & 16000 \\ \hline
\end{tabular}
\centering
\caption{Dataset parameters}
\label{dataset}
\end{table}

Instead of directly using the received time domain complex signal, energy normalization is performed prior to training or inferring. The motivation is three-fold: 1) the impact of energy turns out to be minimal according to simulation results, 2) the modulation structure of the signals can be better exposed without the interference of signal energy, 3) an energy independent model can have a better generalization capability which can work well even if the background noise changes.

\subsection{Customized Two-stage Training}
Two key performance measures for signal detection, namely, $P_f$ and $P_d$, can not be obtained directly from the DL library. Therefore, a callback function in Keras is implemented to compute them for different SNRs at the end of each epoch.

Considering constant false alarm rate (CFAR) detector, a customized two-stage training strategy is designed. In the first stage, early stopping with 6 epochs patience is applied to train the model to convergence. In the second stage, metrics trade-off characteristic is observed that the validation loss and accuracy both keep stable while $P_f$ and $P_d$ at different SNRs varies from epoch to epoch. Therefore, we set a $P_f$ stop interval first, continue from the best model in the first stage and stop training when $P_f$ falls into it. One drawback of DL methods is the lack of precise performance control, applying the two-stage training strategy, we can control detection performance to some extent by adjusting the preset stop interval. The trade-off between control precision and training time is achieved by interval size parameter. A smaller interval attains more precise performance control, but also results in longer training time.

\subsection{Simulation Results}
\label{simulation}
In this section, extensive simulation results are provided to demonstrate the performance of the proposed model.\footnote{For reproducible research, all source codes can be found at https://github.com/EricGJB/DL-based-signal-detection.} Also, the impact of key parameters such as modulation scheme and sample length is investigated.

\subsubsection{Comparison with Different Networks}
\begin{figure}[htbp]
\centering
\includegraphics[width=0.40\textwidth]{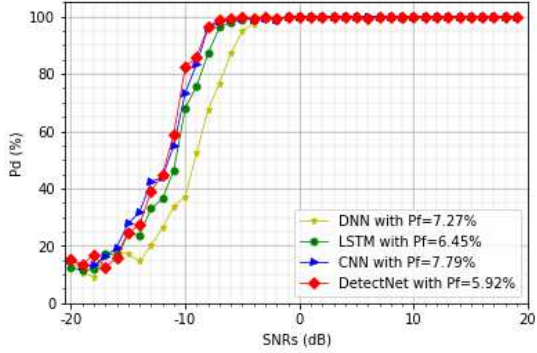}
\caption{Detection performance for various DL models.}
\label{DetectionPerformance}
\end{figure}

Fig. \ref{DetectionPerformance} compares the detection performance of the proposed model with several other popular neural network models on QAM16 signals with sample length of 128. For fair comparison, extensive cross-validation is performed for all models to determine the best hyperparameters. In particular, all models use 0.2 dropout. The DNN consists of four FC layers with 256, 500, 250, 120 neurons respectively, the CNN uses two Conv layers with 60 filters with filter size of 10 and one FC layer with 128 neurons, while the LSTM uses two 128 cells LSTM layers.

It can be observed that DetectNet and CNN achieve better detection performance than DNN and LSTM. In addition, DetectNet attains similar $P_d$ as CNN but outperforms CNN by achieving lower $P_f$. To illustrate the advantage of DetectNet over energy detector, considering the operating point with $P_f=5.92\%$ and $P_d=90\%$. According to Equation \ref{SNR-wall}, the SNR-wall of energy detector is -5.35 dB. From Fig. \ref{DetectionPerformance}, the SNR-wall of DetectNet is -8.5 dB, which is 3.15 dB lower.

\subsubsection{Impact of Modulation Scheme}
\begin{figure}[htbp]
\centering
\includegraphics[width=0.40\textwidth]{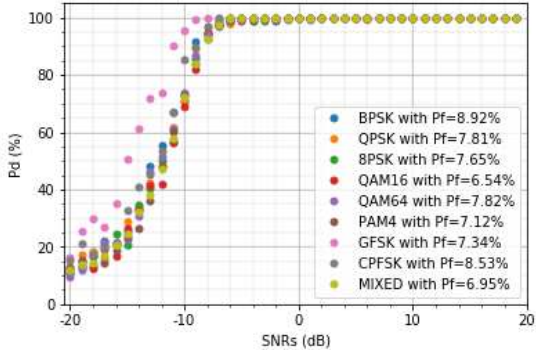}
\caption{Impact of modulation scheme on performance.}
\label{ModulationScheme}
\end{figure}
Fig .\ref{ModulationScheme} illustrates the detection performance of DetectNet over different types of modulation schemes with sample length of 128. As can be observed, the performance for FSK signals, especially GFSK, is better than for PSK and QAM signals. %When $P_f=7.34\%$ and $P_d=90\%$, the SNR-wall of DetectNet over GFSK signals is -12 dB, which is 5.43 dB lower than that of the energy detector.
Moreover, it is surprise to see that the detection performance difference between BPSK, QPSK and 8PSK is rather insignificant, which implies that the performance of DetectNet is insensitive to the modulation order.

\subsubsection{Generalization Ability}
\begin{figure}[htbp]
\centering
\includegraphics[width=0.430\textwidth]{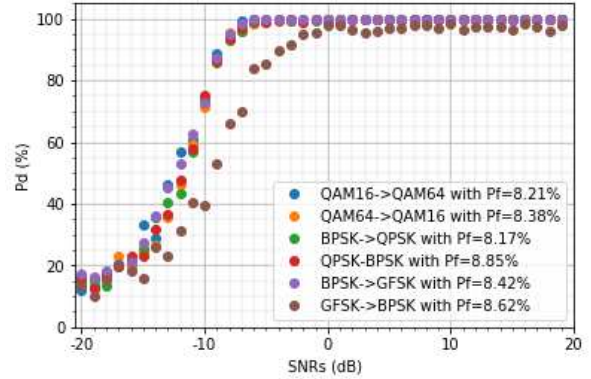}
\caption{Generalization ability to different modulation schemes.}
\label{generalization}
\end{figure}

Fig. \ref{generalization} illustrates the generalization ability of the proposed DetectNet. In particular, we test the detection performance of a well trained network over signals with different modulation schemes from the training signals. Comparing Fig .\ref{ModulationScheme} and Fig. \ref{generalization}, it can be readily observed that as long as the  modulation type is the same, for instance, QAM16 and QAM64, or BPSK and QPSK, the DetectNet provides decent generalization ability. In contrast, if the modulation type is different, for instance, BPSK and GFSK, there is a significant performance deterioration. The reason is rather intuitive, since the DetectNet exploits underlying structural information of modulated signals, which is similar between signals with the same modulation type while differs substantially between signals with different modulation types.

\subsubsection{Impact of Sample Length}
Intuitively, the detection performance of DetectNet improves when the sample length increases due to more available information. Recall that the detection performance of energy detector also improves with longer sample length, it is hence of particular interest to see the performance gap of the two detectors with different sample lengths. Let us consider the operating points with $P_d=90\%$ and set the $P_f$ stop interval as $[7\%,9\%]$ , then the SNR-walls of energy detector and DetectNet on GFSK signals with different sample lengths are summarized in Table \ref{WallImprovement}. It turns out that regardless of the sample length, DetectNet consistently yields around 5dB improvements over the energy detector.
\begin{table}[htbp]
\begin{tabular}{|c|c|c|c|c|}
\hline
$P_f(\%)$ & Sample length & EDW& DLW & Improvement \\ \hline
8.05 & 64 & -3.91 & -9.00 & 5.09 \\ \hline
7.34 & 128 & -5.57  & -11.00 & 5.43\\ \hline
8.45 & 256 & -7.41 & -12.30 & 4.89 \\ \hline
7.73 & 512 & -8.94 & -13.50 & 4.56\\ \hline
7.86 & 1024 & -10.55 & -15.60 & 5.05\\ \hline
\end{tabular}
\centering
\caption{SNR-wall improvement for different sample lengths. EDW and DLW represents for SNR-wall of energy detector and DL based detector respectively. EDW is calculated by Eq .\ref{SNR-wall} with $M \to \infty$. The unit of SNR-wall is dB.}
\label{WallImprovement}
\end{table}

\section{DL BASED COOPERATIVE DETECTION}
Cooperative sensing, which utilizes distributed nodes to work in a collaborative fashion, has been demonstrated to be an efficient means to improve the detection performance. For cooperative sensing, the fusion center makes the final decision based on the hard information from each sensing node. As such, it is not able to exploit the confidence information of the decision of each node. In addition, the priority of different nodes is not used. Motivated by this, in this section, we introduce a DL based cooperative detection system which implicitly exploits these soft information.

\subsection{System Design}
For each sensing node, the DetectNet is employed locally to obtain the probability vector of two hypotheses about primary signal. Then, it is fed into the fusion center for further processing. Unlike in the conventional sensing system where a specific fusion rule is used to combine the hard decision information from the distributed nodes, a neural network consisting of three FC layers is proposed to directly learn the best fusion rule through training.
%\textcolor{blue}{\cite{COML} adopts similar idea while the probability vectors used in \cite{COML} are theoretically derived through energy detection, and in this letter they come from the output of DetectNet, thus are more precise.}

\begin{figure}[htbp]
\centering
\includegraphics[width=0.4\textwidth]{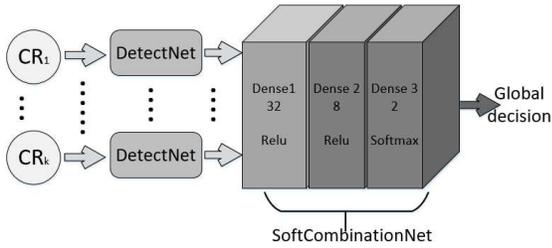}
\caption{DL based cooperative detection system design.}
\label{CooperativeSystem}
\end{figure}

Through extensive cross-validation, the numbers of neurons of each FC layer are given by 32, 8 and 2, respectively. The cooperative detection system model is termed as ``SoftCombinationNet'' and the detailed network architecture is illustrated in Fig. \ref{CooperativeSystem}.

\subsection{Simulation Results}
For simulations, it is assumed that the channel gains between the primary transmitter and $k$ sensing nodes are independently and identically distributed (i.i.d.). Also, experiments are conducted on QAM16 signals with a sample length of 128, and three cooperative systems with 2, 4 and 8 nodes respectively are considered.

Fig. 6 depicts the detection performance of cooperative sensing schemes. For illustration purpose, the Logical-OR (LO) rule is used in the conventional cooperative detection scheme, since it in general yields the highest $P_d$. Comparing the performance of SoftCombinationNet (SCN) and LO, we find that for all three systems, in the practical SNR regime of interests, i.e., where $P_d$ is larger than 90$\%$, SoftCombinationNet achieves almost same $P_d$ as LO, but with a significant reduction in the $P_f$, thereby demonstrating its supriority.

\begin{figure}[htbp]
\centering
\includegraphics[width=0.40\textwidth]{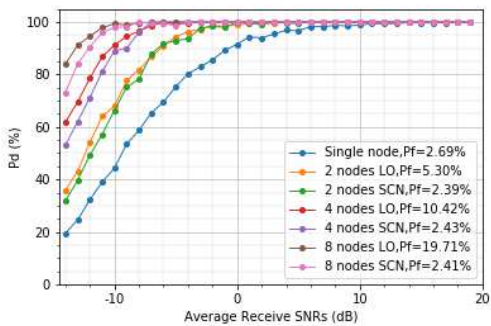}
\caption{Performance gain of DL based cooperative detection system.}
\label{CooperativeResults}
\end{figure}

\section{CONCLUSION}
\label{conclusion}
In this letter, we have proposed a novel DL based signal detector named DetectNet, which exploits the inherent structural information of modulated signals. It was shown that significant performance improvement can be achieved over the conventional energy detector. Also, the DL based detector is insensitive to the modulation order, hence have good generalization ability to similar modulation schemes. Then, a DL based cooperative detection scheme named SoftCombinationNet is proposed to exploit the soft information from distributed sensing nodes, which is shown to achieve high $P_d$ and low $P_f$ simultaneously.

%\nocite{*}
%\bibliographystyle{unsrt}
%\bibliography{references}

\begin{thebibliography}{99}
\bibitem{CR1}
S. Zhang, et al, ``Novel spectrum sensing and access in cognitive radio networks,'' {\em Sci. China Inf. Sci.}, vol. 61, no. 8, 089302, Aug. 2018.

\bibitem{CR2}
J. Zhao, X. Guan, and X. Li, ``Power allocation based on genetic simulated annealing algorithm in cognitive radio networks,'' {\em Chinese journal of Electronics}, vol. 22, no. 1, pp. 177-180, Jan. 2013.

\bibitem{CR3}
J. Zhao, T. Yang, Y. Gong, J. Wang, and L. Fu, ``Power control algorithm of cognitive radio based on non-cooperative game theory,'' {\em China Commun.}, vol. 10, no. 11, pp. 143-154, Nov. 2013.

\bibitem{tandra2008snr}
R. Tandra and A. Sahai, ``SNR walls for signal detection," \emph{IEEE J. Sel. Topics Signal Process}., vol. 2, no.1, pp. 4-17, Feb. 2018.

%\bibitem{alink2011lowering}
%M. S. Oude Alink, et al, ``Lowering the SNR wall for energy detection using cross-correlation," \emph{IEEE Trans. Vel. Technol}., vol. 60, no. 8, pp. 3748-3757, Oct. 2011.

\bibitem{krizhevsky2012imagenet}
A. Krizhevsky, I.Sutskever, and G. E. Hinton, ``Imagenet classification with deep convolutional neural networks," \emph{Proc. Adv. Neural Inf. Process. Syst}., 2012, pp. 1097-1105.

\bibitem{jiang2018artificial}
P. Jiang \emph{et al}. ``Artificial Intelligence-aided OFDM receiver: Design and experimental results," 2018. [Online]. Available: https://arxiv.org/abs/1812.06638

\bibitem{o2017introduction}
X. You, et al, ``AI for 5G: research directions and paradigms," \emph{Sci. China Inf. Sci.}, vol. 62, no. 2, 021301, Feb. 2019.

\bibitem{wang2017deep}
T. Wang, et al, ``Deep learning for wireless physical layer: Opportunities and challenges," \emph{China Commun}., vol. 14, no. 11, pp. 92-111, Nov. 2017.

\bibitem{Azmat}
F. Azmat, Y. Chen, and N. Stocks, ``Analysis of spectrum occupancy using machine learning algorithms," \emph{IEEE Trans. Veh. Technol}., vol. 65, no. 9, pp. 6853-6860, Sep. 2016.

\bibitem{DLSS}
Q. Chen \emph{et al}. ``Deep learning network based spectrum sensing methods for OFDM systems," 2019. [Online]. Available: https://arxiv.org/pdf/1807.09414.pdf

\bibitem{cldnn}
T. N. Sainath, et al, ``Convolutional, long short-term memory, fully connected deep neural networks," in Proc. 2015 IEEE ICASSP, 2015, pp. 4580-4584.

\bibitem{liang2008sensing}
Y. Liang, Y. Zeng, E. C. Y. Peh, and A. T. Hoang, ``Sensing-throughput tradeoff for cognitive radio networks," \emph{IEEE Trans. Wireless Commun}., vol. 7, no. 4, pp. 1326-1337, Apr. 2018.

\bibitem{quan2009optimal}
Z. Quan, S. Cui, A. H. Sayed, and H. V. Poor, ``Optimal multiband joint detection for spectrum sensing in cognitive radio networks," \emph{IEEE Trans. Signal Process}., vol. 57, no. 3, pp. 1128-1140, Mar. 2009.

\bibitem{ED}
S. Atapattu, C. Tellambura and H. Jiang, ``Conventional energy detector," in \emph{Energy Detection for Spectrum Sensing in Cognitive Radio} (Springer Briefs in Computer Science). New York, NY, USA: Springer, 2014, pp. 11-26.

\bibitem{mariani2011effects}
A. Mariani, A. Giorgetti, and M. Chiani, ``Effects of noise power estimation on energy detection for cognitive radio applications," \emph{IEEE Trans. Commun}., vol. 59, no. 12, pp. 3410-3420, Dec. 2011.

\bibitem{mr}
N. E. West and T. J. O'Shea, ``Deep architectures for modulation recognition," in Proc. 2017 IEEE DySPAN, 2017, pp. 1-6.

\bibitem{radioml}
T. J. O'Shea and N. West, ``Radio machine learning dataset generation with gnu radio," in \emph{Proceedings of the GNU Radio Conference}, vol. 1, no. 1, 2016.

%\bibitem{yucek2009survey}
%T. Yucek and H. Arslan, ``A survey of spectrum sensing algorithms for cognitive radio applications," \emph{IEEE Commun. Surveys and Tutorials}., vol. 11, no. 1, pp. 116-130, Quat. 2009.

%\bibitem{COML}
%Y. Lu, P. Zhu, D. Wang and M. Fattouche, ``Machine learning techniques with probability vector for cooperative spectrum sensing in cognitive radio networks," \emph{2016 IEEE Wireless Communications and Networking Conference}, Doha, 2016, pp. 1-6.

\end{thebibliography}

\end{document}